# High-energy topological edge states and strain-induced multiple flat bands in a honeycomb lattice

*Xiaoqin Bai* [1, 2], *Haozhen Tian* [1, 2], *Boris A. Malomed* [4, 5], *Rongcao Yang* [1, 2, *], *Xiaojun Jia* [3, §]

[1] *School of Physics and Electronics Engineering, Shanxi University, Taiyuan 030006, China*

[2] *Shanxi Key Laboratory of Wireless Communication and Detection, Shanxi University, Taiyuan 030006, China*

[3] *State Key Laboratory of Quantum Optics Technologies and Devices, Institute of Opto-Electronics, Shanxi University, Taiyuan 030006, China*

[4] *Department of Physical Electronics, School of Electrical Engineering, Faculty of Engineering, Tel Aviv University, Tel Aviv 69978, Israel*

[5] *Instituto de Alta Investigación, Universidad de Tarapacá, Casilla 7D, Arica, Chile*†

[*] Corresponding author: sxdxyrc@sxu.edu.cn

[§] Corresponding author: jiaxj@sxu.edu.cn

**Abstract** We propose a novel anti-twig edge in the honeycomb lattice (HCL) that supports two symmetric high-energy edge states. It is different from the twig edge supporting the zero-energy flat band. Moreover, multiple flat bands are produced by applying a strain to the HCL with a twig edge or an anti-twig edge, and the suppression or enhancement of the high-energy edge state is observed. Under the edge-parallel stretch strain, the high-energy edge state band merges into the bulk band, suppressing the high-energy edge state, while the zero-energy edge state becomes delocalized. On the other hand, under the action of the edge-parallel compression strain, both the high-energy and zero-energy edge states exhibit strong localization. Pseudo-topological protection of the high-energy edge state is explored too. Finally, by reconstructing the anti-twig edge in the HCL, degenerate flat bands and strain-induced multiple flat bands are produced, and *topologically protected* vacated anti-twig edge states are demonstrated.

**Keywords:** anti-twig edges, high-energy edge states, strain-induced multiple flat bands, pseudo-topological edge, topologically protected edge

## 1. Introduction

Edge states, which can be topologically nontrivial or trivial ones, are widely supported by various lattices. Among them, topologically protected edge states give rise to a unique phenomenon: the conductance remains confined to the material's surface while the bulk remains completely insulating [1,2]. Topological insulators (TIs), a novel state of matter originally introduced in solid-state physics, also exhibit this property, hosting the conductance at the surface, while being an insulator in the bulk. This unique characteristic originates from the existence of topological edge states and renders the topological insulator a subject of much interest in studies of acoustics [3,4], ultracold atoms [5], quantum information [6], and photonics [7-10]. Photonic TIs are typically constructed using spatially periodic artificial materials, with intrinsic structures such as Lieb [11-13], kagome [11,14], honeycomb lattices [9,15-17], and others. Among them, the honeycomb lattice (HCL) serves as a versatile platform for the exploration of topological photonics, which can be tailored to feature zigzag, bearded or armchair edges [18]. These edge geometries are known to support edge states. In particular, the zigzag and bearded edges support topological edge states in the first Brillouin zone (BZ), which are nearly complementary with respect to each other, while the defect-free or isotropic armchair edge does not give rise to topological edge states [18]. The zigzag, bearded and anisotropic armchair edge states have been observed in photonic graphene [18-20]. In 2023, a distinctive edge configuration named the twig edge was proposed, as one complementary to the armchair edge, by engineering an appropriate primitive cell. The twig edge states are characterized by a topologically protected flat band that spans the entire BZ [21].

More recently, the possibility of the formation of topological edge states, and the existence and dynamics of edge-state solitons in the presence of nonlinearity have become the focus in studies of topological effects in HCL photonic systems. Topological edge states and a variety of solitons, including dark, bright, vector and Bragg solitons, have been generated and investigated in spatial inversion-symmetry-broken and time-reversal-symmetry- broken photonic HCLs [22-28]. Dynamical phenomena in edge states, such as Rabi oscillations [29], Bloch oscillations, and Zener tunneling [17,30] have also been reported in photonic HCLs.

Applying a strain to photonic lattices can induce a pseudomagnetic field and the corresponding Landau levels [31,32]. Also, it has been reported that the strain can restore the *PT*-symmetry in complex photonic HCLs, induce deformations in the energy band, unify valley and anomalous Hall effects, and facilitate the creation and destruction of zigzag and bearded edge states [31-35]. In strained photonic HCLs with zigzag and bearded terminations, zero-energy edge states, as well as bright and dark zero-energy edge solitons, have been demonstrated [36]. Additionally, the inhibition and reconstruction of the Zener tunneling have also been realized in strained photonic HCLs with these terminations [30]. Up to date, despite the growing interest in HCLs, their counterparts with a novel twig edge remained unexplored – in particular, new high-energy edge states and multiple flat bands are still undisclosed.

In this work, a novel anti-twig edge in the HCLs is designed, and two symmetric high-energy edge states are obtained. Moreover, multiple flat bands are produced by applying strains to the configured HCLs, and suppression and enhancement of localization of the high-energy edge state are explored under the edge-parallel stretch and compression strains. The localization and topological-like protection of the high-energy edge states are demonstrated too. Furthermore, by reconstructing the configured HCLs, degenerate flat band, multiple flat bands, and topological edge states are investigated, along with the associated topological protection.

## 2. Band structure and edge states produced by the tight-binding approximation

We consider an HCL configuration featuring a specially designed novel edge of the twig type (referred to as an *anti-twig edge*) along the $x$ direction. In contrast to the twig edge, the anti-twig one is formed by incorporating an additional lattice site, thus creating a new type of the boundary feature with a reversed twig structure. The HCL with the anti-twig edge is achieved by selecting the unit cell enclosed by the blue box in Figure 1(a), rather than cutting the bulk along the $x$ or $y$ direction [18]. The considered HCL configuration adopts periodic boundary conditions in the $x$ direction with period $L_x = 3a$ ( $a$ is the lattice constant) and open boundary conditions in the $y$ direction. Consequently, the Bloch momentum $k_x \in [-K_x/2, K_x/2]$ ( $K_x \equiv 2\pi/L_x$ ) is a well-defined quantum number, unlike $k_y$. We choose the super unit cell marked by the shaded rectangle, and calculate the corresponding Bloch Hamiltonian:

$$H = \begin{bmatrix} 0 & \mathcal{H}_{12} & 0 & 0 & 0 & 0 & 0 & 0 \\ \mathcal{H}_{12}^* & 0 & 0 & 0 & \mathcal{H}_{34} & 0 & 0 & 0 \\ 0 & 0 & 0 & \mathcal{H}_{34} & \mathcal{H}_{35} & 0 & 0 & 0 \\ 0 & 0 & \mathcal{H}_{34}^* & 0 & 0 & \mathcal{H}_{35}^* & \mathcal{H}_{12} & 0 \\ 0 & \mathcal{H}_{34}^* & \mathcal{H}_{35}^* & 0 & 0 & H_{12} & 0 & 0 \\ 0 & 0 & 0 & \mathcal{H}_{35} & \mathcal{H}_{12}^* & 0 & 0 & 0 \\ 0 & 0 & 0 & \mathcal{H}_{12}^* & 0 & 0 & 0 & \mathcal{H}_{34} \\ 0 & 0 & 0 & 0 & 0 & 0 & \mathcal{H}_{34}^* & 0 \end{bmatrix}, \quad (1)$$

where $\mathcal{H}_{12} = t_2 \exp\left(-i\mathbf{k}_{\mathbf{edge}} \cdot \mathbf{e}_2\right)$, $\mathcal{H}_{34} = t_3 \exp\left(i\mathbf{k}_{\mathbf{edge}} \cdot \mathbf{e}_3\right)$ and $\mathcal{H}_{35} = t_1 \exp\left(i\mathbf{k}_{\mathbf{edge}} \cdot \mathbf{e}_1\right)$, with $\mathbf{k}_{\mathbf{edge}} = [k_x, 0]$ being the Bloch wave vector, $t_{1,2,3}$ are strengths of the coupling between the nearest-neighbor sites along three different directions, defined by bond vectors $\mathbf{e}_1 = [-a, 0]$, $\mathbf{e}_2 = [a/2, \sqrt{3}a/2]$ , and $\mathbf{e}_3 = [a/2, -\sqrt{3}a/2]$ . The corresponding band structure for $t_1 = t_2 = t_3 = 1$ and $a = 1.4$ is plotted in Figure 1(b). One can clearly see that the proposed HCL supports three edge bands, which are marked by two mutually symmetric red lines and the horizontal green one. The red and green lines are associated with the symmetric edge states which belong, respectively, to the anti-twig and twig edges [see examples of the edge states in Figures 1(d1) and 1(d2)]. Note that the second-order derivative, i.e., the diffraction, of the anti-twig edge states corresponding to the red bands is nonzero [as indicated by the red curve in Figure 1(c)], therefore we designate the corresponding edge state as high-energy edge state; on the contrary, the green band remains completely flat [see the green line in Figure 1(c)], suggesting to designate the corresponding mode as a zero-energy edge state.

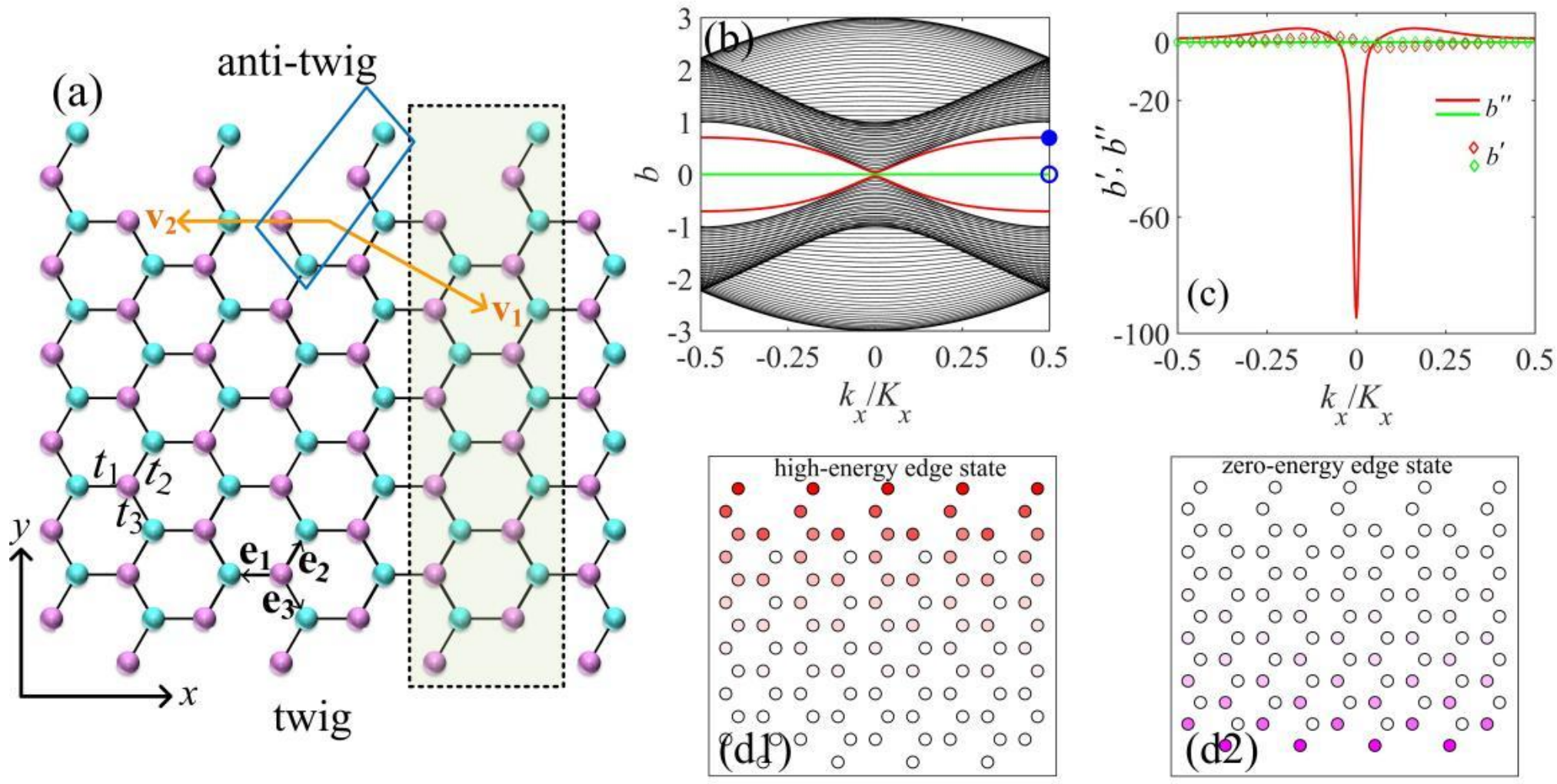

**Figure 1:** (a) Schematic of the HCL with the anti-twig (top) and twig (bottom) edges. $t_{1,2,3}$ represents the coupling strength between the nearest-neighbor sites along three different directions. $\mathbf{e}_1=[-a,0]$, $\mathbf{e}_2=[a/2,\sqrt{3}a/2]$ and $\mathbf{e}_3=[a/2,-\sqrt{3}a/2]$ are bond vectors. The unit cell enclosed by the blue box is defined for anti-twig edge. $\mathbf{v}_1=\left[3a,-\sqrt{3}a\right]$, $\mathbf{v}_2=[-3a,\ 0]$ are the Bravais vectors. The shaded rectangle denotes the super unit cell for the construction of Hamiltonian (1). (b) Band structure corresponding to (a), where the red lines represent two symmetric high-energy edge states and the green line denotes the zero-energy edge state. (c) Velocities $b'$ (hollow rhombuses) and diffraction $b''$ (curves) for the upper red branch and green branch of the bands in (b). (d) High-energy (d1) and zero-energy (d2) edge states at $k_x/K_x=0.5$, which correspond to the blue solid dot and hollow dot in (b), respectively.

We consider an example displayed in Figure 1(d) with a specific Bloch wave vector $k_x/K_x=0.5$, which supports strong localization of the high-energy and zero-energy edge states. We stress that a decrease of the Bloch wavenumber leads to expansion of both the high-energy and zero-energy edge states into the bulk, as shown in Figure 2, which compares the localization of the edge states for $k_x/K_x=0.5$ and $k_x/K_x=0.3$. It is seen that reducing the Bloch wavenumber $k_x$ narrows the separation between the edge bands and the bulk band [see Figures 2(a1) and 2(a2)], resulting in weaker localization of both the high-energy and zero-energy edge states, as seen in Figures 2(b1) and 2(b2), respectively. Moreover, it is evident that the high-energy and zero-energy edge states exhibit exponential localization at their respective anti-twig and twig edges, underscoring the distinct behavior of these edge states.

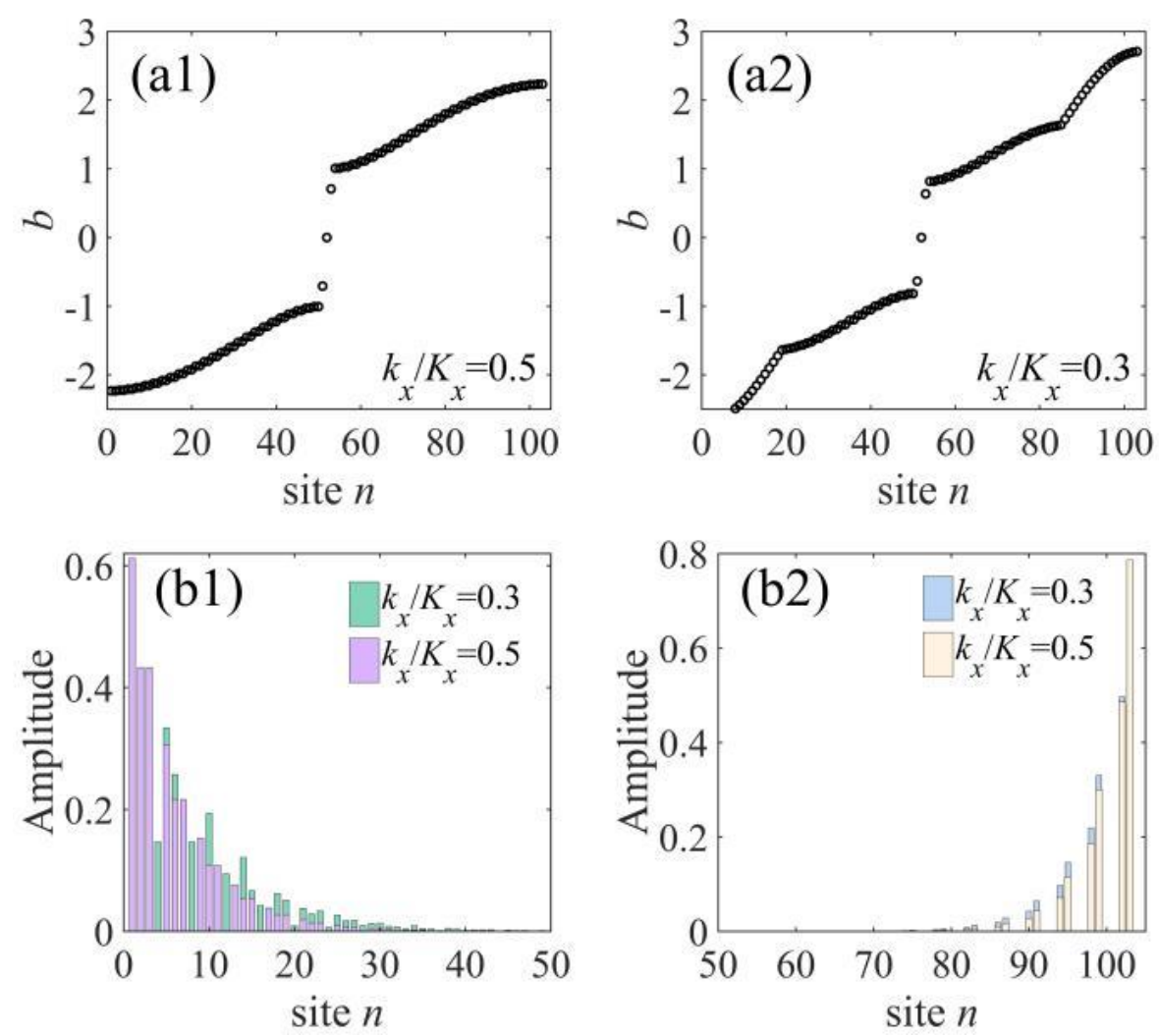

**Figure 2:** (a1, a2) Eigenvalue spectra for $k_x/K_x=0.5$ (a1) and $k_x/K_x=0.3$ (a2) corresponding to Figure 1(b). Panels (b1) and (b2) display, respectively, the high-energy and zero-energy edge states, for $k_x/K_x=0.5$ and $k_x/K_x=0.3$.

## 3. Optical realization of the proposed HCL

In this section, we introduce a feasible physical realization of the proposed HCL, in which the lattice sites can be defined in terms of the refractive-index variation. The photonic lattice can be regarded as an array of waveguides arranged in a honeycomb structure. The desired coupling strength can be established by adjusting the spatial separation between the waveguides. In the photonic HCL, the propagation of light along the $z$ axis is governed by the Schrödinger-like paraxial wave equation for the dimensionless light field $\Psi$:

$$i\frac{\partial \Psi}{\partial z} = -\frac{1}{2}\left(\frac{\partial^2}{\partial x^2}+\frac{\partial^2}{\partial y^2}\right)\Psi - \mathcal{R}(x,y)\Psi \qquad (2)$$

where $x$ and $y$ are the normalized transverse coordinates, $z$ is the propagation distance. $\mathcal{R}(x,y) = p\sum_{m,n}\exp\left[-(x-x_m)^2/d_x^2-(y-y_n)^2/d_y^2\right]$ describes the profiles of the waveguide array as the combination of Gaussian functions, where $(x_m, y_n)$ denotes the positions of the waveguides, $d_x$ and $d_y$ represent the width of a single waveguide, and $p$ is the depth. Such photonic HCL can be fabricated using the femtosecond laser writing method [15,36]. We choose the representative parameters as $p=10$, $d_x=d_y=0.5$ and $a=1.4$, which, respectively, correspond to the refractive-index modulation depth $\sim 10^{-3}$, waveguide width $5\mu\text{m}$ and lattice constant $14\mu\text{m}$, assuming that laser radiation is coupled into the waveguide at the wavelength of 800 nm, with a characteristic transverse size of 10 μm. Accordingly, the lattice with the anti-twig and twig edges, displayed in Figure 1(a), can be emulated optically using the waveguide arrays represented by the potential term $\sim \mathcal{R}(x,y)$, as shown in Figure 3(a). In these waveguide arrays, we investigate the propagation of the high-energy and zero-energy edge states through direct simulation using Eq. (2), and employ an elongated Gaussian excitation condition $\Psi(x,y,z=0) = A_0\exp\left[-x^2/w_x^2-(y-y_0)^2/w_y^2\right]$ with $A_0$ being the amplitude of the excitation, $w_x$ and $w_y$ denoting the width of the excitation, and $y_0$ representing the position of the edges along the $y$ axis. The corresponding propagation dynamics of the high-energy and zero-energy edge states for $k_x/K_x=0.5$ and $k_x/K_x=0.3$ are shown in Figures 3(b) and 3(c). One can easily find that the beams initially break up into the patterns that corresponding to edge states illustrated in Figures 1(d1) and 1(d2). Subsequently, the beams propagate along the edges without obvious scatter into bulk for $k_x/K_x=0.5$, as shown in Figures 3(b1) and 3(b2). However, for $k_x/K_x=0.3$, the reduced localization of both the high-energy and zero-energy edge states implies the weak penetration into bulk, as depicted in Figures 3(c1) and 3(c2). Moreover, it can be observed by comparing Figure 3(c1) with 3(b1) that the high-energy edge state at $k_x/K_x=0.3$ experiences significant diffraction, which is attributed to the non-zero second-order derivative.

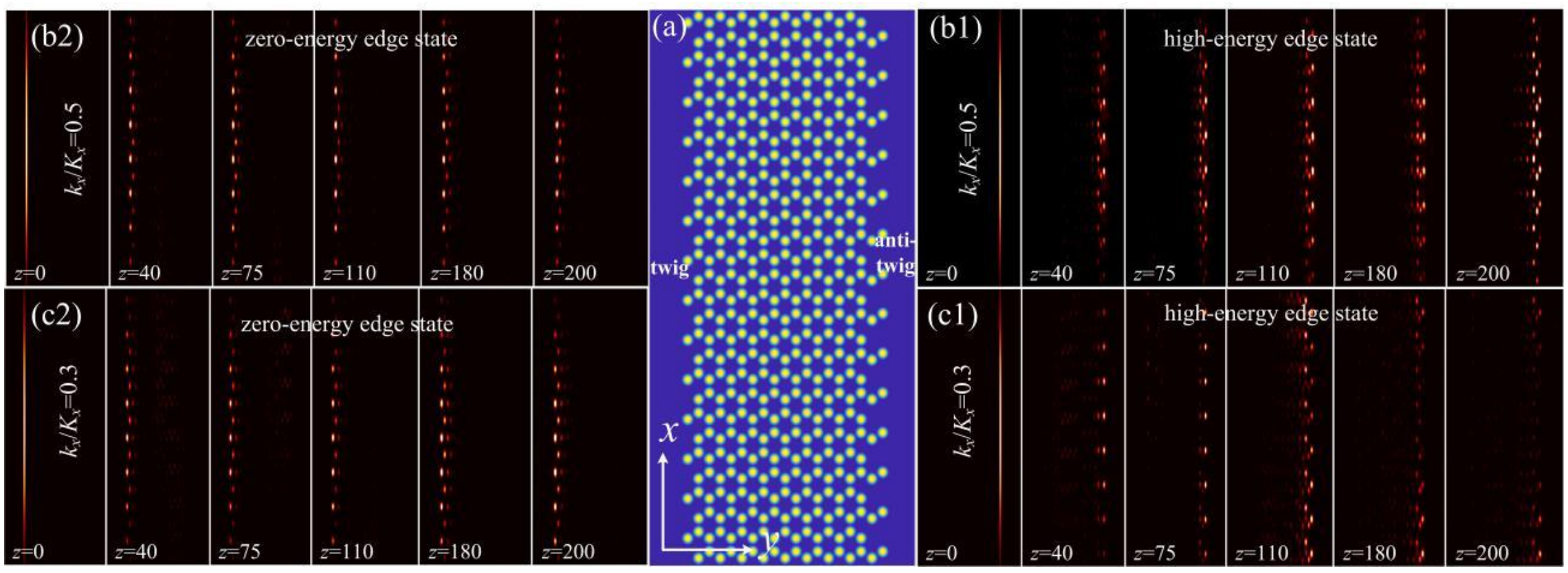

**Figure 3:** (a) Optical realization of the HCL with anti-twig and twig edges. (b, c) Propagation of the high-energy (b1, c1) and zero-energy edge states (b2, c2) at $k_x/K_x=0.5$ (b) and $k_x/K_x=0.3$ (c). Other parameters: $A_0=1$, $w_x=20a/\sqrt{3}$, $w_y=0.5a/\sqrt{3}$.

## 4. Strained photonic HCL

By applying a strain to the proposed HCL, the coupling strength between sites can be efficiently controlled, which can further lead to the deformation of the energy bands. For the transversely stretched HCL (applying an edge-parallel stretch strain), the nearest-neighbor distance between horizontal sites is stretched and the coupling strength is reduced [see Figure 4(a)], i.e. $|\mathbf{e}_1| > |\mathbf{e}_{2,3}|$ and $t_1 < t_{2,3}$. In this case, the transversely stretched HCL can be looked upon as a stack of trivial dimer Su-Schrieffer- Heeger (SSH) chains with stronger intra-cell hopping, as shaded by the gray-dashed region in Figure 4(a), thereby, the edge state is inherently not supported. Figure 4(b) illustrates the band structures of the transversely stretched HCL. It is seen that the upper and lower edge bands tend to merge with the bulk band, and the middle flat band approaches the bulk band, as shown in Figure 4(b1). The decreasing ratio of coupling strength $\delta_s = t_1/t_2$ promotes the eventual disappearance of the upper and lower symmetric edge bands and a narrowing of the separation between the flat band and the bulk band, as presented in Figure 4(b2). This suggests that for the transversely stretched HCL, the modes exhibit notably delocalization, concentrating in both the bulk and edges, as shown in Figures 4(c1) and 4(c2). The wavefunction of the modes shown in Figure 4(d) further indicates the delocalization and the substantial penetration into the bulk. The underlying mechanism for these phenomena is that the stretch strain increases the separation between transverse sites, thereby suppressing the coupling strength. Concurrently, the stretching reduces the lattice confinement, which pushes the high-energy edge band into the bulk. As a result, the energy of the edge states eventually merges with the bulk as the stretch strain intensifies.

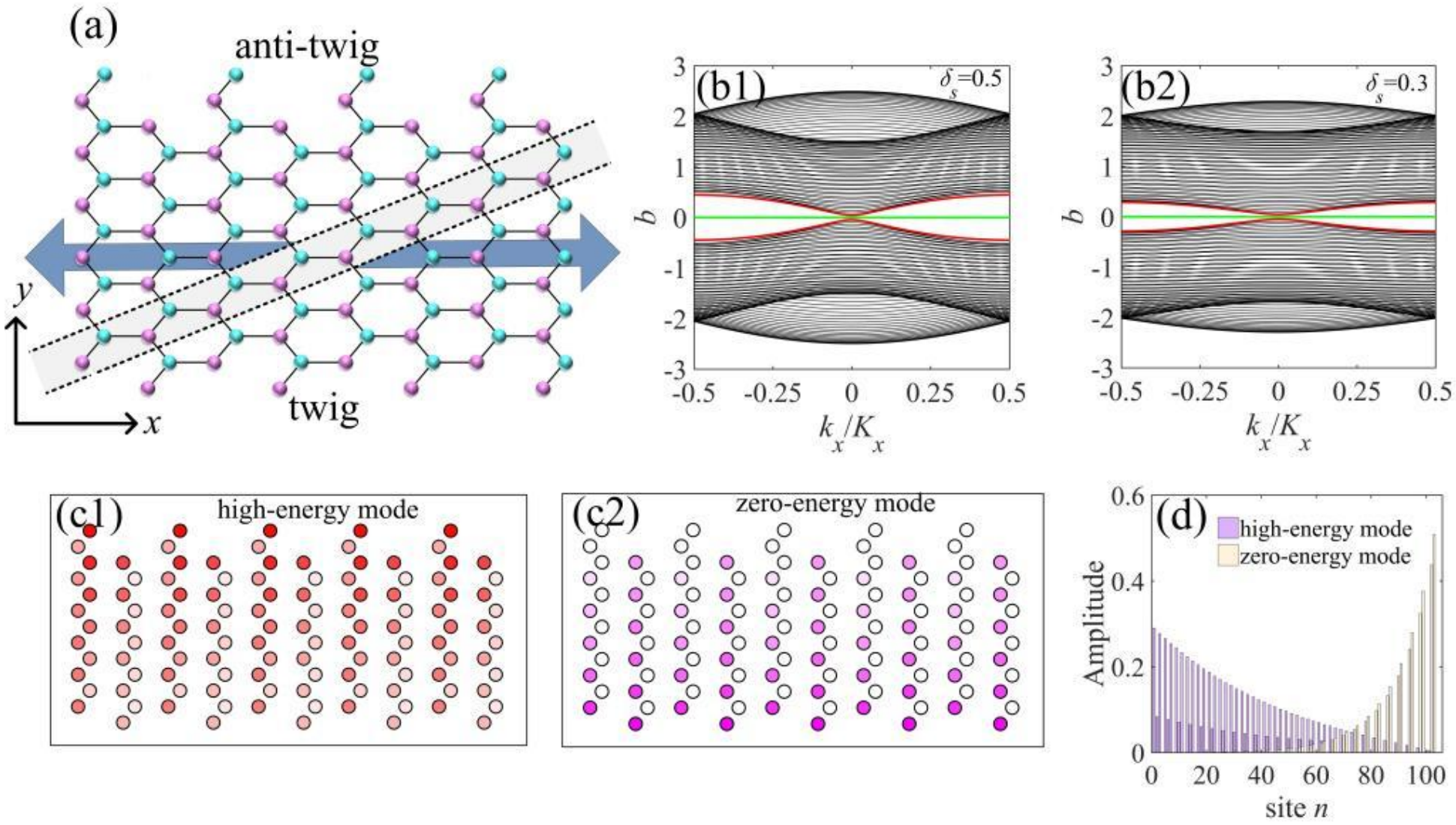

**Figure 4:** (a) Schematic of the transversely stretched HCL with anti-twig and twig edges. Blue arrows represent the direction of stretch strain. (b) Band structures of the transversely stretched HCL corresponding to (a) for $\delta_s = 0.5$ (b1) and $\delta_s = 0.3$ (b2). (c) High-energy (c1) and zero-energy (c2) modes for $k_x/K_x = 0.5$. (d) Wavefunction of the high-energy and zero-energy modes corresponding to (c).

For the transversely compressed HCL (applying an edge-parallel compression strain), the nearest-neighbor horizontal distance between sites is compressed, that is $|\mathbf{e}_1| < |\mathbf{e}_{2,3}|$ and $t_1 > t_{2,3}$. In this case, the strained HCL can be regarded as a stack of SSH chains with weaker intra-cell coupling [36], as highlighted by the gray shading in Figure 5(a). Figures 5(b1) and 5(b2) show the band structures for different ratios of coupling strength $\delta_c = t_2/t_1$. It can be seen that decreasing the ratio $\delta_c$ leads to the disappearance of Dirac point, the opening of band gap, and the emergence of multiple flat bands. This indicates the strong localization of the higher-energy and zero-energy edge states, as shown in Figures 5(c1), 5(c2) and 5(d). These phenomena occur because compression strain decreases the transverse site distance and enhances the hopping integral and coupling strength. Also, the strain-induced anisotropy opens the bandgap, flattens the energy bands, producing flat bands characterized by highly localized states with zero group velocity. It is noted that here the obtained additional non-degenerate edge states are entirely different from that reported in Ref. [37] in which the strained graphene lattice possesses degenerate flat bands.

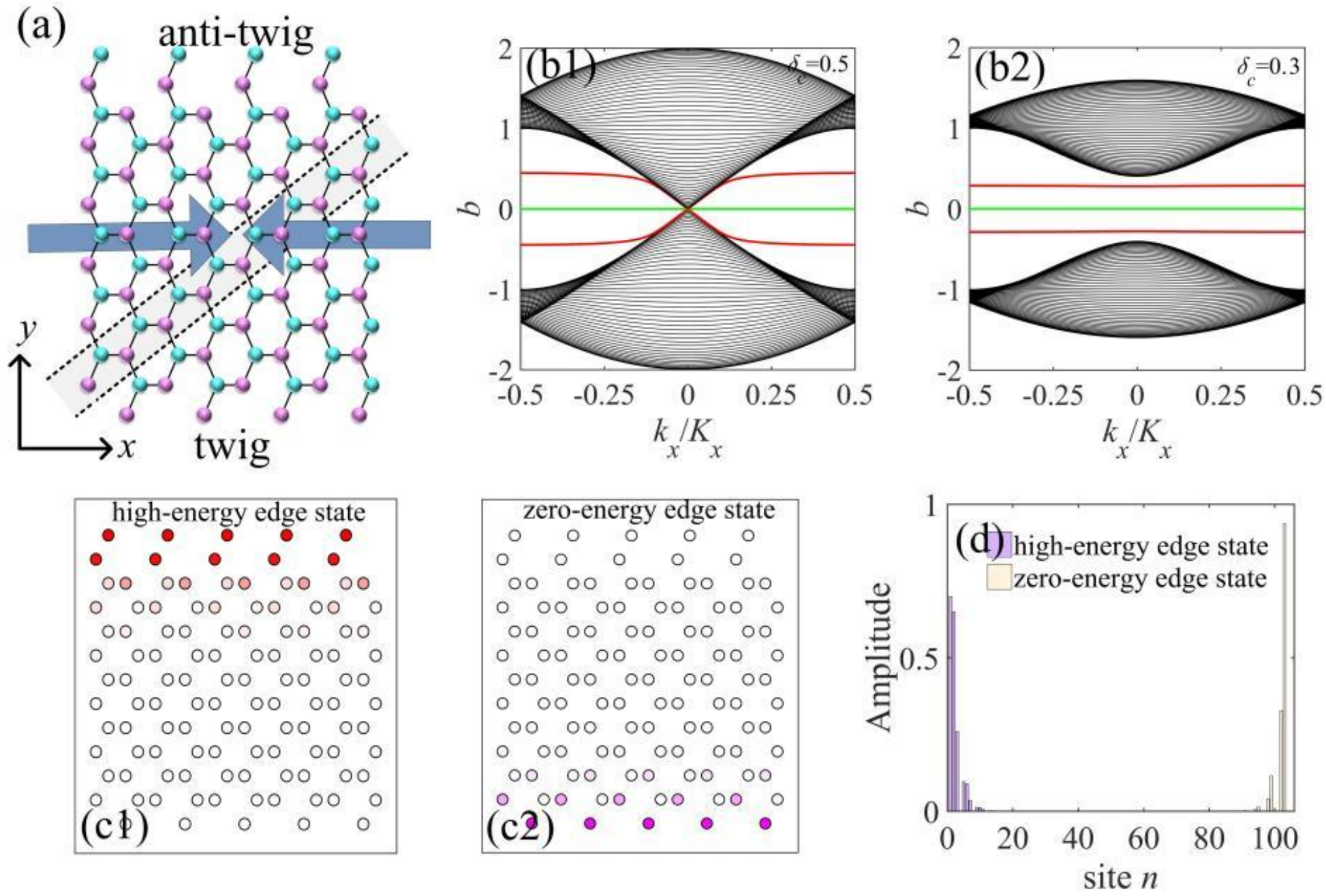

**Figure 5:** (a) Schematic of the transversely compressed HCL with anti-twig and twig edges. Blue arrows represent the direction of compression strain. (b) Band structures of the transversely compressed HCL corresponding to (a) for $\delta_c = 0.5$ (b1) and $\delta_c = 0.3$ (b2). (c) High-energy (c1) and zero-energy (c2) edge states for $k_x/K_x = 0.5$ . (d) Wavefunction of the high-energy and zero-energy edge states corresponding to (c).

To seek conditions for the opening of a bandgap and the emergence of multiple flat bands, we first construct the bulk Hamiltonian by selecting the blue-box unit cell in Figure 1(a), based on the relation between the anti-twig edge and the bulk. The bulk Hamiltonian is

$$H = \begin{bmatrix} 0 & M_k \\ M_k^* & 0 \end{bmatrix} \tag{3}$$

where $M_k$ is $2\times 2$ matrix,

$$M_k = \begin{bmatrix} t_2 + t_3 e^{-i\mathbf{k_{bulk}}\cdot(\mathbf{v_1}+\mathbf{v_2})} & t_1 e^{-i\mathbf{k_{bulk}}\cdot(\mathbf{v_1}+2\mathbf{v_2})} \\ t_1 e^{i\mathbf{k_{bulk}}\cdot(\mathbf{v_1}+\mathbf{v_2})} & t_3 + t_2 e^{i\mathbf{k_{bulk}}\cdot(\mathbf{v_1}+\mathbf{v_2})} \end{bmatrix}, \tag{4}$$

with the Bloch wave vector $\mathbf{k_{bulk}} = \left[k_x, k_y\right]$ and the Bravais vectors $\mathbf{v}_1 = \left[3a, -\sqrt{3}a\right]$ , $\mathbf{v_2} = \left[-3a,\ 0\right]$ . Diagonalizing the bulk Hamiltonian (3), we obtain the eigenvalues

$$b_{1\sim4}\left(k_x,k_y\right) = \pm\left[t_1^2 + 4t_2^2\cos^2\left(\frac{\sqrt{3}}{2}ak_y\right) \pm 4t_1t_2\cos\left(\frac{\sqrt{3}}{2}ak_y\right)\cos\left(\frac{3}{2}ak_x\right)\right]^{\frac{1}{2}}, \tag{5}$$

which represents the band structure of the proposed HCL. The bandgap closes when $b_{2,3} = 0$ at $k_x = 0$ . Thus, the condition for the bandgap opening is

$$\frac{t_1}{2t_2} = \cos\left(\frac{\sqrt{3}}{2}ak_y\right). \tag{6}$$

Equation (6) indicates that the bandgap always closes when $t_1 < 2t_2$ for $-\pi/\sqrt{3}\,a < \left|k_y\right| < \pi/\sqrt{3}\,a$ . In other words, when $t_1 > 2t_2$ , the Dirac point disappears and the bandgap reopens. The band structures are shown in Figures 1(b), 4(b) and 5(b), following the above derivation.

In the optically mimicked lattice, the strained HCL configuration can be realized by precisely manipulating the spatial separation of the waveguides. Figures 6(a) and 6(c), respectively, present the continuous transversely stretched and compressed HCLs corresponding to the configurations depicted in Figures 4(a) and 5(a). In these continuous strained HCLs, the propagation dynamics of the high-energy state and zero-energy state at $k_x/K_x = 0.5$ are shown in Figures 6 (b) and 6(d). The results in Figure 6(b) reveal that in the transversely stretched lattice, both the high-energy mode and zero-energy mode exhibit significant spreading into the bulk during propagating. The characteristic is consistent with that described in Figures 4(c) and 4(d). In contrast, in the transversely compressed lattice, both the high-energy and zero-energy edge states consistently maintain perfect localization at the anti-twig and twig edges during propagating, being characterized by the absence of diffraction, backscattering and energy leakage into bulk, as depicted in Figure 6(d).

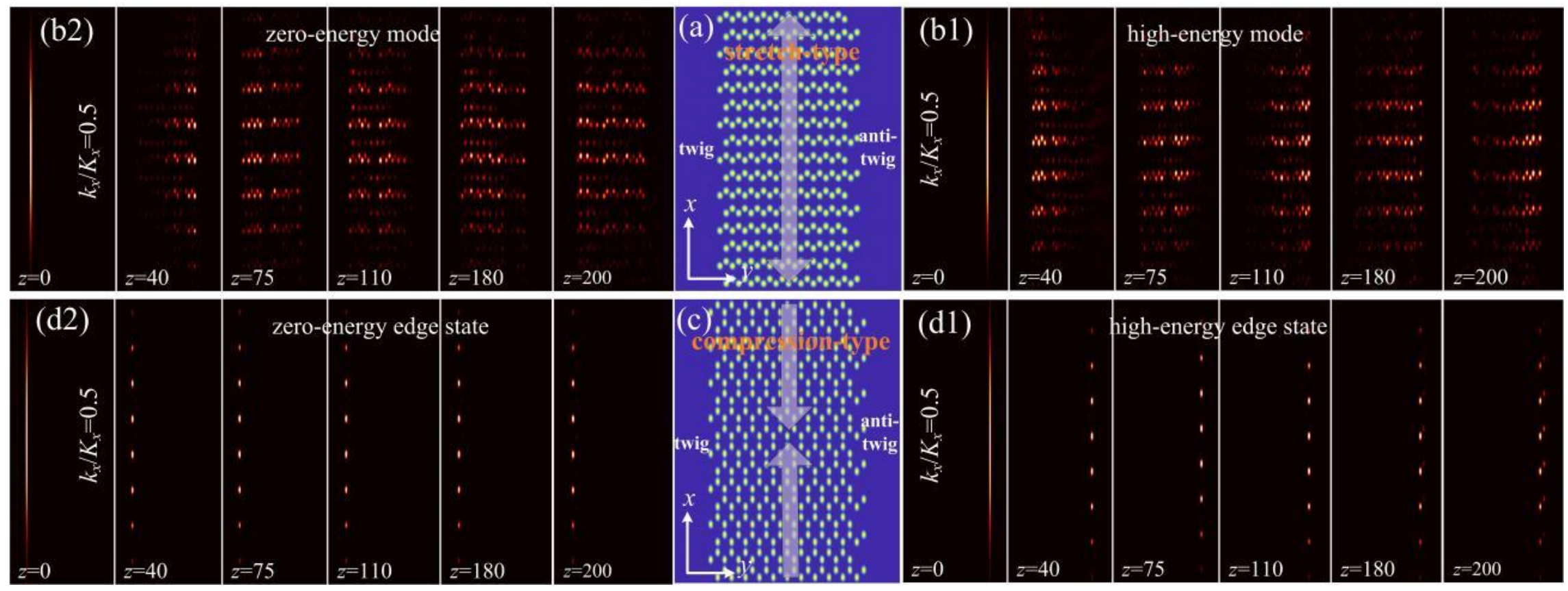

**Figure 6:** (a, c) Optical transversely stretched and compressed HCLs. Purple arrows represent the direction of the strain. (b, d) Propagation of the high-energy state (b1, d1) and zero-energy state (b2, d2) at $k_x/K_x = 0.5$ in the transversely stretched (b) and compressed (d) HCLs. Other parameters are the same as in Figure 3.

## 5. The pseudo-topological nature of the edge states

To explore the topological properties and robust localization of the high-energy edge state, we calculate the winding number as suggested by works [21,37-39]:

$$w = -\frac{1}{2\pi}\int_0^{2\pi} \frac{d}{dk}\arg\left(\det\left(M_k\right)\right)dk \tag{7}$$

in which $k$ is $k_x$ or $k_y$, depending on the edge direction. To calculate the winding number for the anti-twig edge states, the Bloch wave vector $k_x$ is fixed, and the integration is carried out along $k_y$ [37-39]. The resulting winding loops of the high-energy edge state are plotted in the $(\sigma_x, \sigma_y)$ plane in Figures 7(a1-a3). For comparison, the winding loops of the zero-energy edge state [21,37] are also illustrated in Figure 7(a). The origin $O$, marked by a green dot, is encircled by the winding loops of the zero-energy edge state [see the blue loops in Figures 7(a1-a3)], indicating a nonzero winding number and thus its topologically nontrivial nature in the unstrained, transversely stretched and compressed HCLs. The red lines in the figures suggest that the high-energy edge states at the anti-twig edge are topologically trivial. However, strong localization of the high-energy edge states at the anti-twig edge is manifested even in a defective HCL shown in Figure 7(b). The propagation dynamics of the high-energy and zero-energy edge states in the transversely compressed lattice with defects are displayed in Figure 7(c). It can be observed that the high-energy edge state localizes strongly at the lattice edges with defects in the course of the propagation, resembling the topological protection behavior of the zero-energy edge state. Therefore, we characterize such localization behavior of the high-energy edge state as *pseudo-topological protection*. Furthermore, it is noteworthy that the persistent localization of the edge states near the defects is attributed to zero velocity of the edge states at $k_x/K_x = 0.5$ .

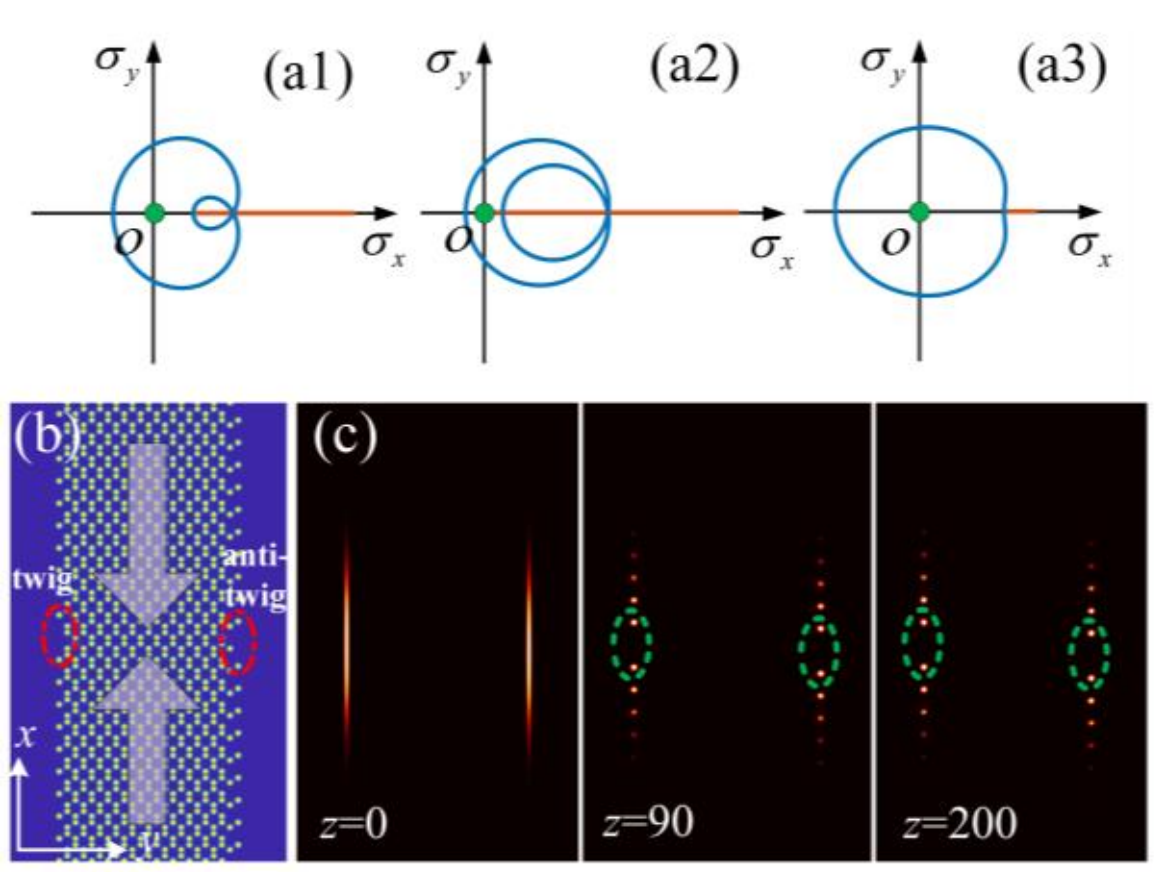

**Figure 7:** (a) Winding loops of the high-energy (red lines) and zero-energy (blue curves) edge states for $k_x/K_x = 0.5$ in unstrained (a1), transversely stretched (a2) and compressed (a3) HCLs. (b) A transversely compressed HCL with two respective defects at anti-twig and twig edges, marked by the red circles. (c) Propagation of the edge states in the transversely compressed lattice shown in (b).

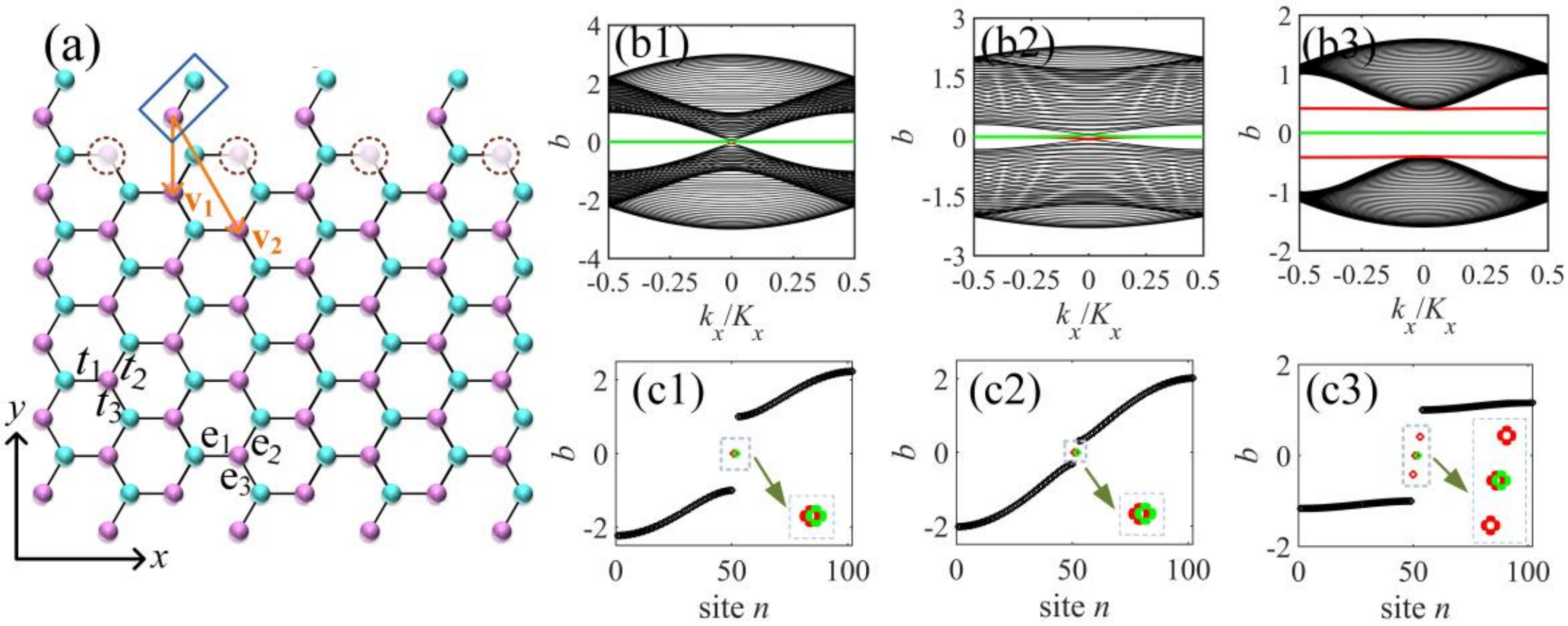

**Figure 8:** (a) The schematic of the HCL with the topologically protected vacated anti-twig (top) and twig (bottom) edges. The unit cell enclosed by the blue box pertains to the vacated anti-twig edge. $\mathbf{v}_1 = \left[0, -\sqrt{3}a\right]$, $\mathbf{v}_2 = \left[3a/2, -3\sqrt{3}a/2\right]$ are the Bravais vectors. (b) Band structures of unstrained (b1), transversely stretched (b2), and compressed (b3) HCLs, where the middle lines represent two degenerate flat bands. (c) Eigenvalue spectra for $k_x/K_x = 0.5$ corresponding to (b).

## 6. Topologically protected vacated anti-twig edge states

To produce true topological edge states, we reconstruct the anti-twig edge pattern by removing the sites enclosed by the dashed circles in Figure 8(a). The so reconstructed anti-twig edge is referred to as a *vacated anti-twig edge*, to distinguish it from the anti-twig edge discussed above. In contrast to the pseudo-topological anti-twig edge shown in Figures 1(a), the vacated anti-twig edge is a *topologically protected one*, as demonstrated below. The band structures of the unstrained, transversely stretched, and compressed HCLs with a vacated anti-twig edge are shown in Figures 8(b1-b3), and the corresponding eigenvalue spectra for $k_x/K_x = 0.5$ are displayed in Figures 8(c1-c3). It is seen that flat bands at the edges appear and the transverse compression strain also leads to multiple flat bands. Note that, unlike the cases of the pseudo-topological anti-twig edge shown in Figures 1(b), 4(b) and 5(b), degenerate flat bands are observed, as marked in Figures 8(c1-c3). Moreover, the energy of the degenerate edge states at the vacated anti-twig edge is localized on the blue sublattices, as depicted in Figure 9(a). Sharp exponential localization is clearly observed in both the unstrained and transversely compressed HCLs [Figures 9(a1) and 9(a3)], in contrast to the weak localization in the stretched lattice [Figure 9(a2)].

To demonstrate the topological nature of the edge states, the unit cell, marked by the blue box in Figure 8(a), is chosen to describe the reconstructed HCL with the vacated anti-twig edge. The corresponding bulk Hamiltonian is written as

$$H = \begin{bmatrix} 0 & Q_k \\ Q_k^* & 0 \end{bmatrix},$$

$$Q_k = t_2 + t_1 e^{i\mathbf{k}\cdot(-2\mathbf{v}_1 + \mathbf{v}_2)} + t_3 e^{-i\mathbf{k}\cdot\mathbf{v}_1} \tag{8}$$

where $\mathbf{k} = \left[k_x, k_y\right]$ is the Bloch wave vector, $\mathbf{v}_1 = \left[0, -\sqrt{3}a\right]$ and $\mathbf{v}_2 = \left[3a/2, -3\sqrt{3}a/2\right]$ are Bravais vectors. The winding loops of the edge states at the vacated anti-twig edge are calculated as per Eqs. (7) and (8). As shown in Figure 9(b), the circle around the origin $O$ (marked by the blue dots) indicates that the edge states are topologically nontrivial according to the bulk-boundary correspondence, with winding loops identical to those at the twig edge. Furthermore, the topological protection of the edge states is examined in the transversely compressed HCL with a defect at the vacated anti-twig edge [see Figure 9(c)]. Figure 9(d) displays the propagation of the edge state, which passes the defect without diffraction or backscattering, confirming the topological protection of the anti-twig edge state.

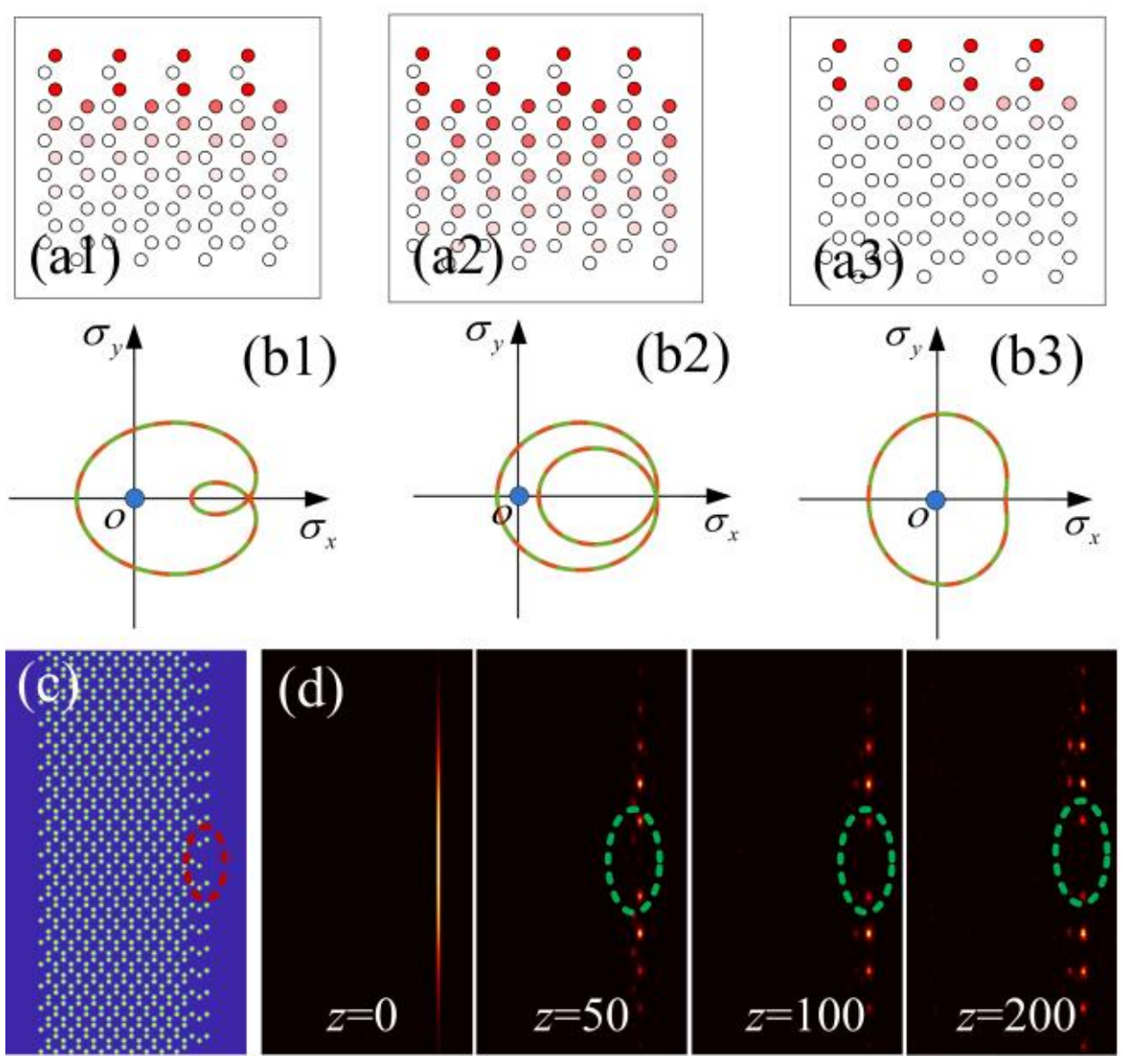

**Figure 9:** (a) Degenerate edge states at the topologically protected vacated anti-twig edge in unstrained (a1), transversely stretched (a2) and compressed (a3) HCLs. (b) Winding loops at the vacated anti-twig edge (red solid curves) and the twig edge (green dashed curves) for $k_x/K_x = 0.5$ in unstrained (b1), transversely stretched (b2) and compressed (b3) HCLs. (c) Transversely compressed HCL with a defect (red circle) at the vacated anti-twig edge. (d) Propagation of the edge state at the edge shown in (c).

## 7. Conclusion

In summary, we have investigated the high-energy edge state in the HCL with a novel anti-twig edge. It is found that the localization of the high-energy edge state depends on the Bloch momenta. By applying strains, we demonstrated the suppression and enhancement of localization of both the high-energy and zero-energy edge states in the HCL. In the transversely stretched HCL, the bands of high-energy and zero-energy edge states undergo progressive disappearance and significant coupling into bulk, respectively, resulting in the delocalization of these states. In the transversely compressed HCL, multiple flat bands appear, and both the high-energy and zero-energy edge states exhibit profound localization at the edges. Employing an optical continuous model, the propagation dynamics of the high-energy and zero-energy edge states are simulated. The topological property of the high-energy edge state is explored by winding loops, and the pseudo-topological protection of the edge state is observed in the transversely compressed HCL. Furthermore, degenerate flat bands and strain-induced multiple flat bands are achieved through reconstructing the anti-twig edge into its topologically protected counterpart. The topological nature and protection of the corresponding edge states are demonstrated. Our research may expand the geometrical construction of topological insulators and underscore the distinct responses to mechanical strain across different lattice types.

**Research Funding** This work was supported by the National Natural Science Foundation of China (Grant No. 62575165, 62305199), the Natural Science Foundation of Shanxi Province (Grant No. 202203021221016) and Shanxi graduate Education Innovation Program (Grant No. 2024YZ06).

**Author Contributions Xiaoqin Bai:** Writing - original draft, Methodology, Formal analysis, Data curation. **Haozhen Tian:** Writing - review and editing, Software. **Boris A. Malomed:** Conceptualization, Formal analysis, Writing - review and editing. **Rongcao Yang:** Supervision, Resources, Methodology, Funding acquisition, Conceptualization. **Xiaojun Jia:** Conceptualization, Resources, Supervision.

**Conflict of Interest** Authors state no conflicts of interest.

**Data Availability Statement** The data that supports the findings of this study are available from the corresponding author upon reasonable request.